\documentclass[10pt,conference]{IEEEtran}
\IEEEoverridecommandlockouts

\ifCLASSINFOpdf
\else
\fi

\usepackage{graphicx}
\usepackage{booktabs}
\usepackage{pifont}
\usepackage{url}
\usepackage{enumitem}

\newlist{customitemize}{itemize}{1}
\setlist[customitemize,1]{
  leftmargin=*, 
  label={\raisebox{-0.2ex}{\scalebox{1.4}{\textbullet}}}, 
  align=left 
}

\usepackage{algorithm}
\usepackage{algpseudocode}
\usepackage{amsmath}

\usepackage{hyperref}
\hypersetup{
	linkcolor=red,
	urlcolor=blue,
	citecolor=green,
}

\usepackage{bbding}
\usepackage{multirow}

\usepackage{tabularx}  

\usepackage{tcolorbox}
\newtcolorbox{mybox}[2][]{
  colback=gray!10,            
  colframe=black,             
  fonttitle=\bfseries\large,        
  coltitle=white,             
  colbacktitle=black,         
  title=#2,                   
  #1                          
}

\usepackage{pgfplots}
\pgfplotsset{compat=1.18} 

\begin{document}

\title{DistillGuard: Malicious NPM Package Detection and API Attack Chain Analysis via Static Graph and LLM Distillation}




\author{
\IEEEauthorblockN{
Siyuan Pang\IEEEauthorrefmark{2}\IEEEauthorrefmark{3},
Yepeng Yao\IEEEauthorrefmark{2}\IEEEauthorrefmark{3}\IEEEauthorrefmark{1}\thanks{\IEEEauthorrefmark{1} Corresponding author: Yepeng Yao},
Zhengwei Jiang\IEEEauthorrefmark{2}\IEEEauthorrefmark{3}\IEEEauthorrefmark{1}\thanks{\IEEEauthorrefmark{1} Corresponding author: Zhengwei Jiang},
Zijing Fan\IEEEauthorrefmark{2},
Baoxu Liu\IEEEauthorrefmark{2}\IEEEauthorrefmark{3}
} 

\IEEEauthorblockA{\IEEEauthorrefmark{2}\textit{Institute of Information Engineering, Chinese Academy of Sciences, Beijing, China}} 
\IEEEauthorblockA{\IEEEauthorrefmark{3}\textit{School of Cyber Security, University of Chinese Academy of Sciences, Beijing, China}}

\IEEEauthorblockA{\normalsize{\{pangsiyuan, yaoyepeng, jiangzhengwei, fanzijing, liubaoxu\}@iie.ac.cn}}
}

\maketitle

\begin{abstract}
The Node.js ecosystem heavily relies on NPM packages, and software supply chain attacks targeting malicious NPM packages are rampant. Malicious code primarily triggers during package installation, import, and runtime. Traditional static analysis fails to understand code semantics; machine learning-based methods rely on feature extraction, which suffers from concept drift; existing LLM solutions suffer from high invocation costs, high data security risks, and poor performance. To overcome these limitations, we propose DistillGuard, a lightweight malicious NPM package detection framework that combines static graph semantic analysis and LLM knowledge distillation. DistillGuard first acquires multi-granular features through three static analysis modules. Then, it leverages online LLM to distill high-quality security knowledge and structured labels. Finally, it uses LoRa to efficiently fine-tune the open-source Qwen3-8B model to support offline deployment. Experiments show that DistillGuard achieves an accuracy of 95.3\%, a precision of 99.4\%, and an F1 score of 93.8\%, outperforming state-of-the-art tools, improving the F1-score by 11.1–30.0 percentage points over the baselines. Our empirical research further reveals the stages of malicious attacks and the distribution of malicious behaviors. We also summarized eight typical API attack chains for malicious activities, providing practical insights for NPM supply chain security defense.

\end{abstract}


\IEEEpeerreviewmaketitle

\section{Introduction} \label{sec:introduction}
Open source software (OSS) has become the infrastructure for modern web and backend software development. As the core package repository of the Node.js ecosystem, NPM hosts millions of public packages, supporting the iterative development and deployment of countless web applications, server-side programs, and frontend projects. However, the high openness and low barrier to entry of the NPM ecosystem also make it a primary target for software supply chain attacks. Attackers typically employ techniques such as domain name hijacking, package hijacking, and malicious code injection to compromise legitimate packages or release counterfeit malicious packages. These malicious packages trigger harmful behaviors during package installation, module import, and project runtime, leading to serious security incidents including remote command execution, reverse shell penetration, sensitive host information leakage, and credential theft. The frequent outbreaks of NPM supply chain threats seriously threaten the security of enterprise business systems and personal user devices, posing an urgent challenge to open source supply chain security governance.

Traditional static analysis tools rely on manually written matching rules for sensitive APIs and dangerous keywords. However, these methods \cite{cousot1977abstract} \cite{tripp2009taj} lack code semantic understanding capabilities, have poor generalization ability against obfuscated and variant malicious code, and generate a large number of false positives due to overly rigid rule matching. Machine learning-based detection methods \cite{chen2016xgboost} \cite{ke2017lightgbm} \cite{breiman2001random} \cite{freund1997decision} rely on manual feature extraction to learn malicious code patterns, but they are severely limited by feature engineering quality and concept drift issues. These methods struggle to capture deep logical relationships and chain-like attack behaviors within the code.

Recently, LLM \cite{liu2024deepseek} \cite{team2025kimi} has been applied to code security analysis. However, current LLM-based NPM malicious packet detection schemes still suffer from several significant practical limitations. Online general purpose large scale models \cite{du2022glm} face high invocation costs, unavoidable data security risks from external data transmission, and insufficient detection performance targeting domain-specific attack logic in NPM. Furthermore, existing LLM methods directly input raw code for inference without mining the structured static features of the packet, failing to combine explicit code structural risks with implicit security semantics, which limits their accuracy and robustness. Additionally, the lack of high-quality domain-specific fine-tuning datasets further hinders the performance improvement of lightweight offline detection models.

To overcome these limitations, we propose DistillGuard, a lightweight malicious NPM packet detection framework that innovatively integrates static graph semantic analysis and LLM knowledge distillation \cite{hinton2015distilling}. Unlike traditional detection methods that rely on single-dimensional features or simple semantic reasoning, DistillGuard constructs multi-granularity risk features from package configuration rules, dependency imports, and function call graphs through a dedicated static analysis module. Building upon this, we utilize online LLM distillation to extract high-quality NPM security domain knowledge, transforming the malicious logic implicit in the code structure into structured regulatory labels. Finally, we employ an efficient LoRA parameter fine-tuning strategy \cite{hu2022lora} \cite{ouyang2022training} to optimize the open-source Qwen3-8B model \cite{qwen3}, enabling this lightweight model to inherit expert-level security reasoning capabilities and support fully offline, low-cost deployment. Experimental results demonstrate that DistillGuard achieves superior detection performance, with an accuracy of 95.3\%, precision of 99.4\%, and an F1 score of 93.8\%, comprehensively outperforming state-of-the-art static analysis, machine learning, and LLM-based detection tools. A series of ablation experiments validate the complementarity and necessity of all static analysis modules within the proposed framework.

Furthermore, this paper conducts a large-scale, systematic empirical study of malicious NPM packages. We quantitatively analyzed and revealed the distribution of various malicious behaviors across three triggering scenarios: installation, import, and runtime. Simultaneously, we categorized all malicious samples by attack type with fine granularity, systematically analyzing the distribution characteristics of malicious behaviors such as information leakage, command execution, malicious downloads, and credential theft. Based on this, we further refined and summarized eight typical malicious API attack chains and their corresponding high-risk dependencies, providing practical and feasible defense guidance for NPM supply chain security.

\textbf{Contributions.} The contributions of our paper are as follows: 

$\bullet$ We propose a novel, lightweight malicious NPM package detection framework, DistillGuard, which combines static graph semantic analysis and LLM knowledge distillation to effectively address the shortcomings of traditional methods.

$\bullet$ We construct a large-scale dataset of benign NPM packages.

$\bullet$ We conduct extensive experiments to validate DistillGuard's superior detection performance and module effectiveness, achieving state-of-the-art detection results.

$\bullet$ We perform a comprehensive empirical analysis of malicious NPM packages, systematically revealing the distribution characteristics of attack stages and malicious behaviors, and summarizing eight typical API attack chains, providing valuable insights for NPM supply chain security defense.


\section{Preliminary} \label{sec:motivation}
\subsection{Background} 

NPM serves as the core package management tool within the Node.js ecosystem; however, its open-source supply chain is frequently plagued by threats from malicious packages. Malicious code typically triggers during three distinct phases: installation, importation, and runtime.

The installation phase initiates automatically when the `npm install' command is executed. During this process, lifecycle scripts—such as `preinstall', `install', and `postinstall' configured within the `package.json' file are executed sequentially. Attackers frequently embed malicious commands and execution logic within these scripts, allowing the code to run directly during the package installation process; this remains the most prevalent attack vector currently employed by malicious NPM packages.

The importation phase occurs when developers introduce modules using `import' or `require' statements. The `main' field within `package.json' points to the module's entry point file, and the initialization code contained within that file executes automatically upon loading. Attackers can implant malicious logic at this juncture, leveraging the standard module loading workflow to execute their attacks—a method that often goes unnoticed by users.

The runtime phase spans the entire lifecycle of a business project. Malicious code in this category typically employs conditional trigger mechanisms, activating only when specific inputs, environmental conditions, or events are met. While this approach significantly increases the difficulty of tracing and remediating the malicious code, its actual execution probability is relatively lower, as its activation is contingent upon the fulfillment of these specific preconditions.

Traditional static analysis techniques rely on manually crafted rules; consequently, they suffer from limited generalization capabilities and struggle to defend against attacks involving code variants. Although general-purpose large language models possess the capacity to comprehend code semantics, they often lack the specialized adaptations required for the specific domain of NPM security. To address these limitations, this paper presents a research study on the detection of malicious NPM packages, integrating static analysis techniques with model distillation and fine-tuning methodologies.

\subsection{Case Study}

\begin{figure}
    \centering
    \includegraphics[width=0.9\linewidth]{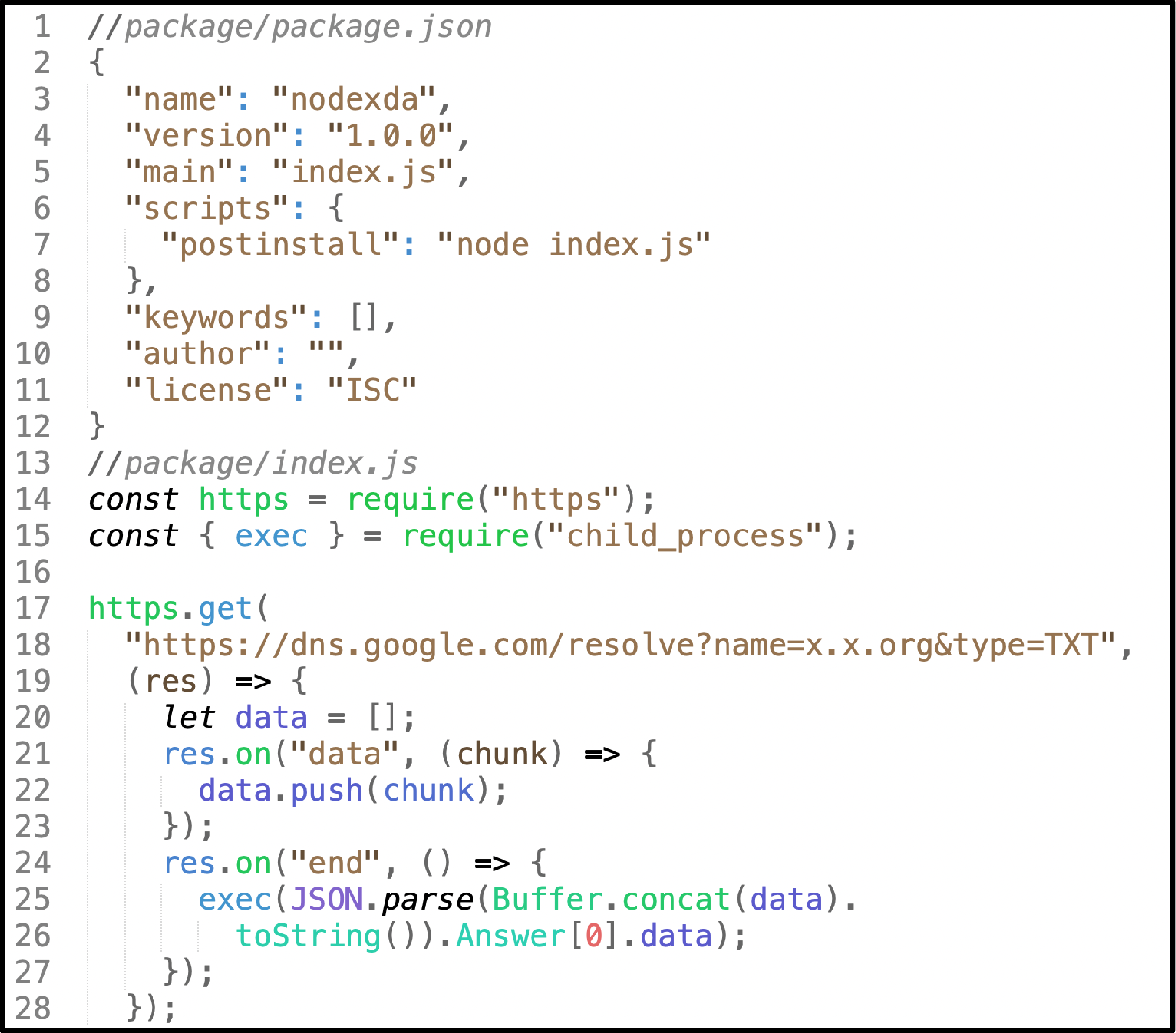}
    \caption{Code Snippet of package `nodexda'}
    \label{case_study_1}
\end{figure}

\textbf{Case Study \#1:} Figure \ref{case_study_1} is a typical malicious NPM package named "nodexda," which executes a classic supply chain attack involving remote command execution during the installation phase.

Within the package's configuration file, `package.json', the attacker maliciously configured the `postinstall' lifecycle script. This configuration ensures that immediately upon the completion of the package installation process, the script automatically invokes and executes the core malicious code contained in `index.js'. When a user executes `npm install' to install this dependency, NPM automatically triggers this script in the background; the entire process requires no manual intervention from the user, making it a highly common and stealthy attack vector during the installation stage. Inside the core code file, `index.js', the malicious package actively imports two high-risk built-in modules: the `https' module for network requests and the `child\_process' module for executing local system commands. These two modules provide the foundational capabilities required for subsequent malicious activities involving remote control and local command execution. Leveraging these modules, the sample constructs a complete malicious execution chain, utilizing the `https.get' and `child\_process.exec' interfaces in sequence to carry out the attack. Specifically, the malicious code actively initiates an HTTPS request to the remote domain `dkoqwdkoqk.duckdns.org', attempting to retrieve the DNS TXT records associated with that domain. The code receives the network response data in segments, concatenates them, and then performs JSON parsing on the complete response content. It extracts data from a pre-defined field within the parsed result and treats this content—fetched directly from the external server—as a system command, which is then executed on the local host via the `exec' function.

This behavior constitutes a standard Command and Control (C2) attack chain, realizing a malicious logic wherein instructions are fetched from a remote server and arbitrary system commands are dynamically executed locally. The entire attack workflow operates fully automatically without issuing any alerts to the user; furthermore, the attacker can dynamically alter the commands being executed simply by modifying the content on the remote server, demonstrating extreme stealth and flexibility. Once successfully executed, this attack can lead to a series of severe security consequences, including remote host compromise, sensitive data leakage, and the unauthorized alteration of system privileges.

\begin{figure}
    \centering
    \includegraphics[width=0.9\linewidth]{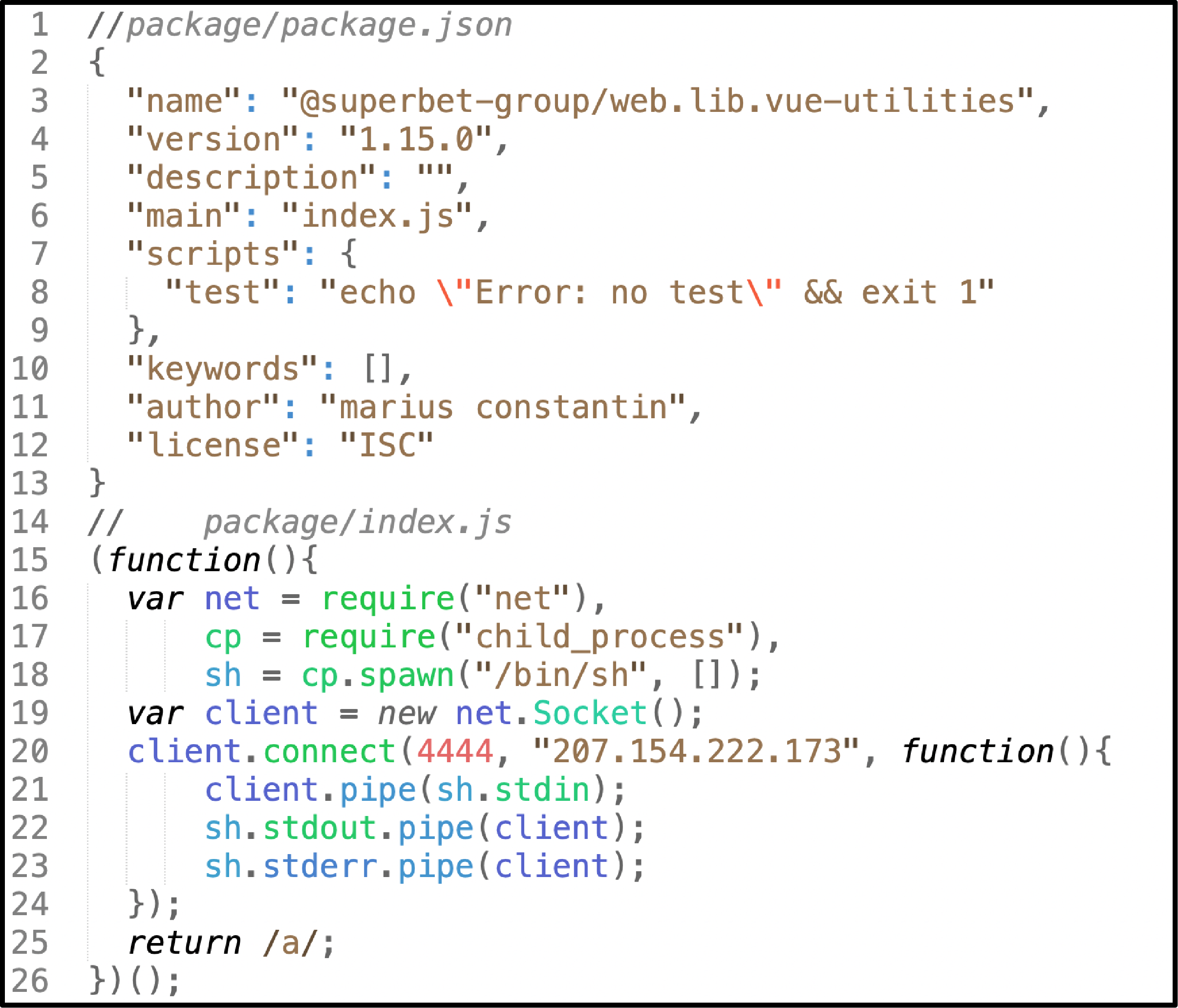}
    \caption{Code Snippet of package `@superbet-group/web.lib.vue-utilities'}
    \label{case_study_2}
\end{figure}

\textbf{Case Study \#2:} This malicious code in Figure \ref{case_study_2} examines a malicious NPM package—named `@superbet-group/web.lib.vue-utilities' —that masquerades as a legitimate utility library. Designed to resemble a Vue-related toolkit, the package is highly deceptive. Unlike samples that rely on scripts triggered during the installation phase, this specimen executes its malicious logic directly when the module is imported; it represents a classic example of an `import-phase' attack.

Within its `package.json' configuration file, the package lists standard package details, author information, and open-source licensing terms. Crucially, it omits high-risk lifecycle scripts—such as `postinstall'—presenting no obvious malicious characteristics on the surface, thereby evading basic detection rules. The `main' field points to `index.js' as the entry point; consequently, the malicious code is executed immediately whenever the module is incorporated into a project via `import' or `require'. The core logic, contained in `index.js', is encapsulated within a self-executing anonymous function that runs automatically upon module loading. The code imports two high-risk built-in modules—`net' and `child\_process'—used for establishing network connections and executing system processes, respectively. The sample first utilizes the `child\_process' module to launch an interactive `/bin/sh' terminal, thereby gaining command-line execution privileges on the host system. It then employs the `net' module to create a socket connection, actively initiating a connection to a remote server located at `207.154.222.173' on port `4444'.

Once the connection is established, the malicious code pipes the terminal's standard input, standard output, and standard error streams to the remote socket, facilitating a bidirectional data pass-through between the host terminal and the attacker's server. This behavior constitutes a classic `reverse shell' attack, enabling the attacker to directly control the victimized host from the remote server—executing arbitrary system commands, reading files, modifying permissions, and deploying persistent malware. 

\begin{figure*}
    \centering
    \includegraphics[width=0.9\linewidth]{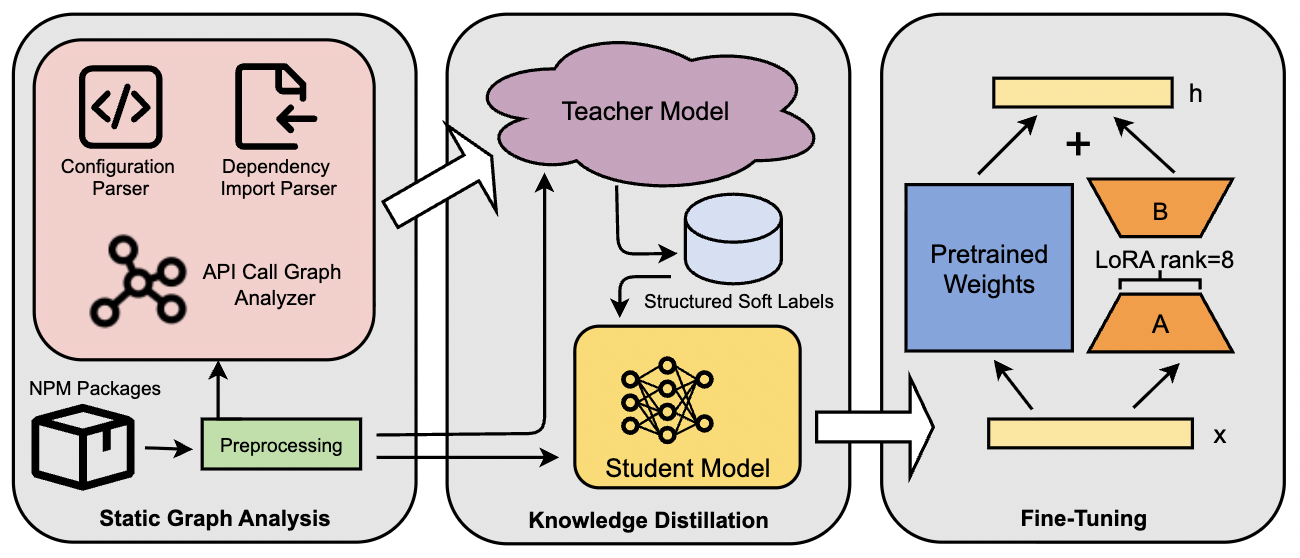}
    \caption{\textbf{The overall architecture of the DistillGuard detection framework first extracts multidimensional features, including static graphs. Then, it uses knowledge distillation based on online teacher LLM to encode the implicit malicious logic in the static features into structured secure soft labels. Finally, it performs efficient fine-tuning based on LoRA on a student model with a small number of parameters to achieve offline, high-precision, and data-secure malicious NPM packet detection.}}
    \label{system}
\end{figure*}

\section{OUR APPROACH} \label{sec:methodology}


\subsection{Overview}
We propose a novel malicious NPM package detection framework called \textbf{DistillGuard}, which combines static graph semantic analysis and online large model knowledge distillation. Unlike traditional detection methods that rely on single static rules or shallow code feature learning, DistillGuard innovatively integrates multi dimensional static structural features and high level security domain knowledge distilled from online LLMs. The entire process consists of three progressive stages: static graph feature extraction, knowledge distillation, and efficient parameter fine-tuning of a lightweight LLM. This framework ultimately constructs an offline secure, lightweight, high precision, and interpretable malicious npm package detection model, effectively addressing the problems of insufficient semantic understanding in traditional static detection and the insecurity and high cost of online large model data.

\subsection{Static Graph Analysis}

The first phase aims to perform a comprehensive static structure analysis on the decompressed original NPM package, extracting multi-granular static features to provide soft labels and structured auxiliary evidence for subsequent knowledge distillation. Unlike simple code text extraction, DistillGuard has designed three dedicated static analysis components to uncover risk clues from package configuration, dependency imports, and function call logic, without executing any malicious code, thus ensuring the security of static analysis.

\subsubsection{Package.json Configuration Parser}

\begin{figure}
    \centering
    \includegraphics[width=\linewidth]{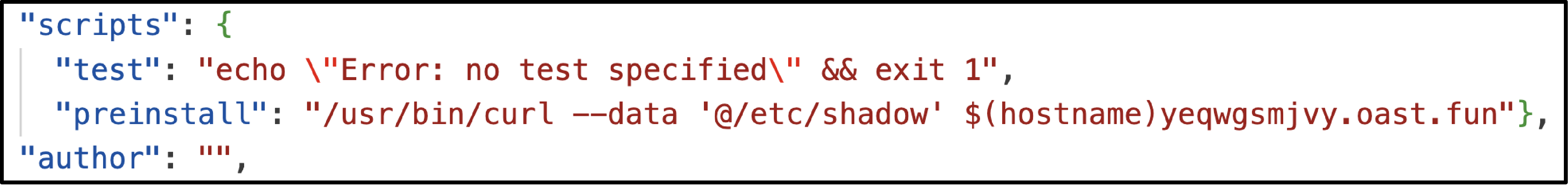}
    \caption{The installation hook script in package.json}
    \label{mal_packagejson}
\end{figure}

As shown in Figure \ref{mal_packagejson}, malicious NPM packages often exploit installation hook scripts to trigger malicious behavior during package installation, which is one of the most common attack methods in supply chain attacks. This component specifically parses the `package.json' configuration file of each NPM package, focusing on scanning dangerous lifecycle hook fields such as `preinstall', `postinstall', and `preuninstall' , as well as other custom execution scripts. The component records all executable script commands defined in the package configuration, filters out abnormal automated execution logic that does not conform to normal package development standards, and retains high-risk configuration characteristics as key evidence.

\subsubsection{Dependency Import Parser}

This component, based on Abstract Syntax Tree (AST) parsing technology, traverses all JavaScript source files of an NPM package, accurately identifying all module import behaviors, such as `require()' and `import', and recording all dependent modules called by the package. It pays particular attention to sensitive system-level modules such as those involved in network communication, file I/O, system information collection, and command execution. The extracted dependency import features reflect the package's potential attack capabilities from the perspective of module dependencies, constituting an important dimension in the characterization of malicious behavior.

\subsubsection{API Call Graph Analyzer}

\begin{figure}
    \centering
    \includegraphics[width=\linewidth]{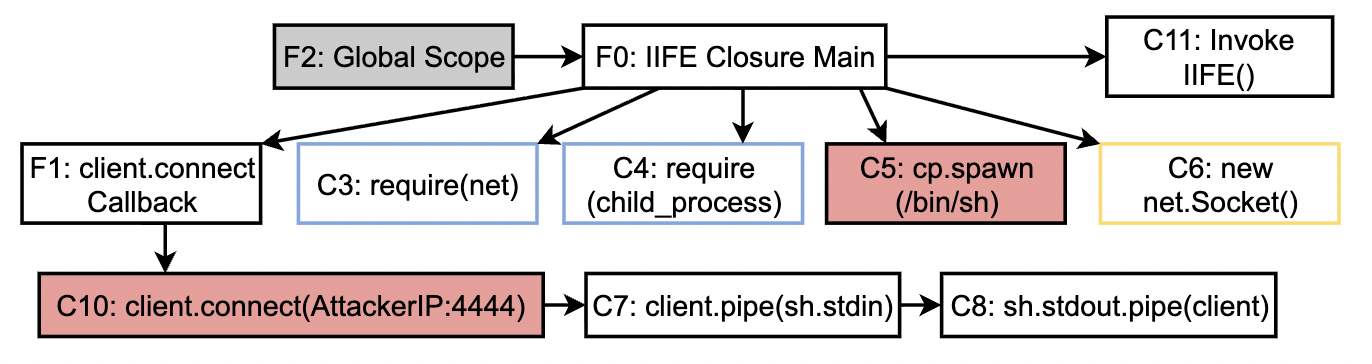}
    \caption{The call graph generated by package `@superbet-group/web.lib.vue-utilities'}
    \label{fig:call_graph}
\end{figure}

To accurately uncover the behavioral logic and execution path of malicious code, this paper employs the JavaScript static analysis tool Jelly \footnote{\url{https://www.npmjs.com/package/@cs-au-dk/jelly}} and completes customized secondary development to construct fine grained function call graphs for each NPM package, as show in Figure \ref{fig:call_graph}. Unlike traditional coarse grained code feature extraction schemes, the modified tool can deeply traverse all function call relationships, analyze the execution sequence of core APIs, and fully reconstruct the call chain of sensitive system functions. Based on the generated call graphs, this component can effectively identify hidden malicious logic that is difficult to detect using rule-based detection methods, including multi level API chain calls and conditionally triggered malicious behaviors. This will better provide analytical features for online large-scale models. After completing static parsing, we use the structured function call graph as an important component of the input prompt, along with other parsing results, and feed it into the online large scale model to support the next stage of knowledge distillation.

\subsection{Knowledge Distillation}

The static graph features extracted in the first stage only retain graph structure and rule matching information, lacking high-level security semantic understanding and malicious behavior reasoning capabilities. To bridge the gap between low-level static features and high-level security judgment, this paper employs the powerful online foundation model GPT-5 \cite{singh2025openai} with a large parameter set for professional security knowledge distillation, transforming the implicit malicious logic involved in graph structure and rule matching into explicit structured supervisory labels.

In this stage, we construct a unified model input for each NPM package. This input integrates three types of information: the concatenated source code of the NPM package, dangerous hook information parsed from `package.json' , a list of sensitive dependencies, and a structured function call graph generated by Jelly. We design only a basic system instruction, commanding the large model to act as an npm package security analysis expert, constraining the LLM to analyze input samples from a supply chain security detection perspective.

Through in-depth reasoning on code semantics, configuration risks, and call logic, GPT-5 outputs standardized structured detection results, including five core dimensions: binary malicious judgment, detection confidence score, malicious API call chain sequence, high-risk dependency modules, and detailed malicious behavior analysis. The distillation results clearly summarize core attack behaviors such as host information gathering, network data theft, domain tunneling, and environment evasion, forming high quality, expert-level supervisory labels. This process transforms the implicit security domain knowledge of online LLMs into labeled data, fundamentally solving the problem of the lack of high quality fine tuned samples in existing malware detection tasks.

\subsection{Fine-Tuning}

\begin{table}[]
\small
\centering
\caption{Classification of Various Malicious Types}
\label{tab:malicious_types_classification}
\begin{tabularx}{\linewidth}{@{}c|X@{}}
\toprule
\textbf{Malicious Category} & \textbf{Description} \\ 
\midrule
Command-Execution & Execute operating system-level shell commands.              \\
\midrule
Code-Obfuscation         & Hiding malicious logic and evading static analysis.         \\
\midrule
Info-Exfiltration          & Collect host information and send it to the attacker's server.              \\
\midrule
Malicious-Download         & Download and execute remote malicious payload.           \\  
\midrule
Credential-Theft        & Stealing SSH keys, kubeconfig, passwords, tokens, etc.      \\
\midrule
DNS-Tunneling          & Data leaks are concealed through DNS queries.      \\
\midrule
Resource-Abuse(Mining)  & Background mining consumes the host's computing power.   \\
\midrule
Persistence  & Injecting code to maintain long-term control.      \\
\bottomrule
\end{tabularx}%
\end{table}

While online pedestal models possess superior secure inference capabilities, they also suffer from insurmountable limitations, such as the inability to deploy locally, high inference costs, and uncontrollable data security and prediction logic. This paper uses a high-quality dataset extracted in the second stage as supervision and employs the LoRA (Low-Rank Adaptive) method to efficiently fine-tune the parameters of the open-source Qwen3-8B model.

We construct an end-to-end instruction tuning dataset. Its input is the complete concatenated source code of an NPM package, and its output is structured security detection knowledge extracted from GPT-5. Unlike traditional binary classification datasets, our dataset contains multi-dimensional, fine-grained supervision information, including confidence scores, malicious API tracking, dependency risk assessment, and behavioral explanations. This enables the open-source model to learn discriminative malicious features and interpretable secure inference logic. In addition, we have categorized malicious behaviors into eight types according to the mainstream attack methods of malicious API chains, as shown in the Table \ref{tab:malicious_types_classification}. We will conduct a detailed empirical analysis of the malicious packets in the future.

During fine-tuning, we freeze the backbone network parameters of Qwen3-8B and only update the low-rank matrix parameters of LoRA. This significantly reduces training memory consumption and avoids overfitting caused by full parameter tuning. After fine-tuning, the lightweight Qwen3-8B model successfully inherited the professional security detection knowledge of the online LLM, enabling it to independently complete end-to-end malicious NPM packet detection, including risk assessment, confidence evaluation, malicious feature localization, and behavioral analysis. The final DistillGuard framework achieves the combined advantages of static graph structure feature extraction, high-level security knowledge reasoning, and lightweight local deployment.

\section{Evaluation} \label{sec:evaluation}
To evaluate DistillGuard, we investigate the following three research questions:

\begin{customitemize}[leftmargin=2em]
    \item \textbf{RQ1:} How was our DistillGuard trained, and what were the results?
\end{customitemize}

\begin{customitemize}[leftmargin=2em]
    \item \textbf{RQ2:} How effective is our tool DistillGuard in detection on the dataset?
\end{customitemize}

\begin{customitemize}[leftmargin=2em]
    \item \textbf{RQ3:} How do the different components of our tool contribute?
\end{customitemize}

\begin{customitemize}[leftmargin=2em]
    \item \textbf{RQ4:} When do malicious packets launch their attacks? What types of malicious behaviors are involved? What is the API attack chain?
\end{customitemize}

\subsection{Experimental Dataset}

\begin{table}[]
\large
\centering
\caption{Statistics of the constructed dataset}
\label{dataset}
\begin{tabular}{@{}l|cc@{}}
\toprule
\multicolumn{1}{l|}{\textbf{Dataset}} & \textbf{\#Malicious} & \textbf{\#Benign}\\
\midrule
Guo et al.\cite{guo2026understanding}         & 4,308  &-     \\
Maloss \cite{Duan2020maloss}          & 567      & -   \\
Ourwork                                   & -      &10,306  \\
\midrule
Total                                     & 4875  &10,306 \\

\bottomrule
\end{tabular}%
\end{table}

\textbf{Malware Datasets.} Given that the detection capabilities and evaluation efficacy of relevant methods depend heavily on large-scale, high-quality datasets, we constructed the evaluation benchmark for this study based on a reliable and manually annotated dataset; this dataset is derived from a highly prestigious and authoritative research paper published by Duan et al. \cite{Duan2020maloss} and Guo et al.\cite{guo2026understanding}Furthermore, we collected a dataset of malware samples provided by the professional security firm Datadog. Detailed statistics regarding these datasets are presented in Table \ref{dataset}.

\textbf{Benign Dataset.} Based on download volume and popularity metrics, this study collected 20,306 software packages from the official NPM repository. We prioritized packages with high download counts and high popularity; such packages possess extensive user bases and benefit from community oversight mechanisms, rendering the probability of malicious code injection extremely low and thereby effectively ensuring the purity of the samples. Furthermore, these popular packages exhibit high code complexity and broad functional coverage, making them more representative of real-world application scenarios. For the experiments, we consistently selected the latest version of each package; these versions have not only undergone comprehensive security maintenance but also reflect current mainstream coding standards. Finally, we performed cross-verification against major security databases to ensure that the benign dataset contains no known malicious packages.

\textbf{Fine-tuning Datasets.}
For the model fine-tuning stage, we selected a subset consisting of 2,000 benign packages and 1,000 malicious packages from the aforementioned full datasets. Alongside the original source code of these selected packages, we also incorporated the intermediate soft labels generated via static graph analysis and knowledge distillation as supplementary input features. The combined dataset configuration, with a total of 3,000 samples and a benign-to-malicious ratio of 2:1, is empirically determined to be well-suited for fine-tuning the 8B parameter large language model adopted in this work. This sample scale and class proportion can effectively balance training efficiency and model generalization ability, while fully enabling the model to learn discriminative patterns between benign and malicious NPM packages.

\subsection{RQ1: Training Results.}





\begin{figure}[htbp]
\centering
\begin{tikzpicture}
\begin{axis}[
    width=0.9\linewidth,
    height=5cm,
    xlabel={Training Step},
    ylabel={Loss Value},
    grid=major,
    grid style={gray!25,dashed},
    legend pos=north east,
    thick
]
\addplot[blue,smooth] table[x=Step,y=Loss,col sep=comma]{./table/train_loss.csv};
\end{axis}
\end{tikzpicture}
\caption{Loss curve during training}
\label{fig:train_loss}
\end{figure}
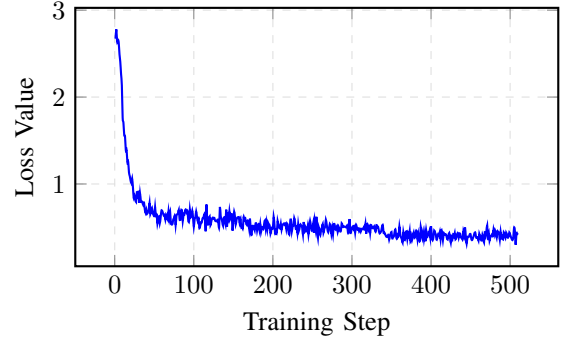

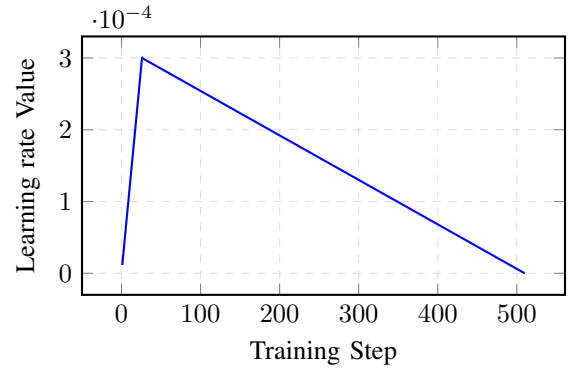
\begin{figure}[htbp]
\centering
\begin{tikzpicture}
\begin{axis}[
    width=0.9\linewidth,
    height=5cm,
    xlabel={Training Step},
    ylabel={Learning rate Value},
    grid=major,
    grid style={gray!25,dashed},
    legend pos=north east,
    thick
]
\addplot[blue,smooth] table[x=Step,y=Value,col sep=comma]{./table/train_lr.csv};
\end{axis}
\end{tikzpicture}
\caption{Changes in learning rate during training}
\label{fig:train_lr}
\end{figure}


    
    


This research constructs a specialized instruction-tuning dataset specifically for the task of detecting malicious npm packages, organizing training samples using a standardized human-AI interaction instruction format. Each sample consists of three components: an instruction, an input, and an output. The instruction component establishes the model's persona as a professional malicious npm package detector, explicitly requiring it to perform a comprehensive assessment based on the source code. The input component concatenates the `package.json' configuration file with the project's main source code, and also incorporates intermediate features extracted through static analysis. The output component is not manually annotated; instead, it leverages GPT-5 to perform knowledge distillation and generate standardized, structured results. This standardized sample format guides the large language model in establishing an end-to-end mapping between raw code and security detection conclusions, thereby enhancing the model's ability to capture and generalize malicious code features.

\textbf{Training Parameter Configuration.} The experiments utilized the Qwen3-8B large language model as the base model, employing the LoRA method to implement parameter-efficient fine-tuning. This approach significantly reduced computational overhead while fully preserving the general capabilities of the pre-trained model. The experimental setup involved 3 full passes over the dataset; a linear decay strategy was selected for the learning rate schedule, accompanied by a learning rate warm-up ratio of 0.05. Given that the concatenated text comprising npm package source code and configuration files often resulted in substantial length, the maximum sequence length per sample was set to 32,768 to accommodate the entire input content. The remaining key training parameters were configured as follows: a training batch size of 16, and an initial learning rate set to \(3\times10^{-4}\). The LoRA-specific parameters were configured with a rank of 8, a scaling factor of 16, and a dropout rate of 0.1. Additionally, a weight decay coefficient of 0.01 was applied to constrain the model parameters via L2 regularization, thereby mitigating the issue of overfitting. The entire fine-tuning task, from initiation to completion, took a total of 3 hours and 6 minutes, with the overall training efficiency meeting expectations.

\textbf{Training Results.} Regarding the training loss, the loss in Figure \ref{fig:train_loss} value peaked at 2.7786 at step 2 of training. As the number of training steps increased, it declined rapidly, eventually reaching a minimum value of 0.3012 at step 507. During the early stages of training, the loss dropped significantly, indicating that the model quickly learned the distinguishing features between benign and malicious packets. In the mid to late stages of training, the loss curve gradually flattened out as the model's feature learning entered a saturation phase; throughout the entire process, no training divergence issues such as loss oscillation or sustained increases were observed.

The learning rate int Figure \ref{fig:train_lr} reached its maximum value of 0.0003 at step 26 and decreased to 0 at step 510, strictly adhering to the preset linear decay schedule. The warm up phase saw a steady increase in the learning rate, while the formal training phase involved a uniform reduction in the parameter update step size; this effectively ensured rapid parameter convergence during the early stages of training and fine-grained parameter optimization during the later stages.

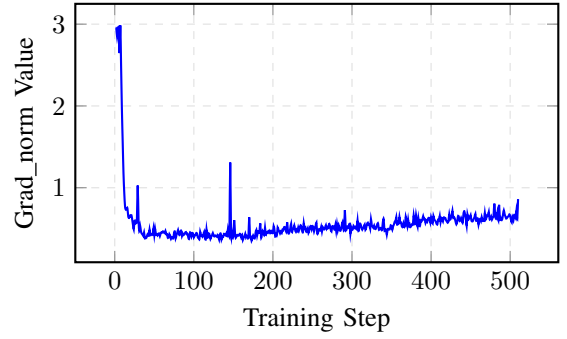
\begin{figure}[htbp]
\centering
\begin{tikzpicture}
\begin{axis}[
    width=0.9\linewidth,
    height=5cm,
    xlabel={Training Step},
    ylabel={Grad\_norm Value},
    grid=major,
    grid style={gray!25,dashed},
    legend pos=north east,
    thick
]
\addplot[blue,smooth] table[x=Step,y=Value,col sep=comma]{./table/train_grad_norm.csv};
\end{axis}
\end{tikzpicture}
\caption{Gradient norm change curve during training}
\label{fig:train_grad_norm}
\end{figure}

In terms of gradient norms in Figure \ref{fig:train_grad_norm}, the maximum gradient norm value 2.9849 occurred at step 7, while the minimum value 0.3528 occurred at step 173. The gradient norm decreased rapidly during the initial stages of training and subsequently remained at a low level with only minor fluctuations; this confirms that the model's gradient propagation functioned normally, with no occurrence of typical training failures such as vanishing or exploding gradients.

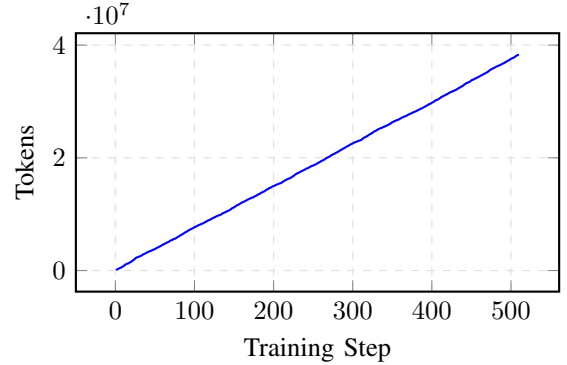
\begin{figure}[htbp]
\centering
\begin{tikzpicture}
\begin{axis}[
    width=0.9\linewidth,
    height=5cm,
    xlabel={Training Step},
    ylabel={Tokens},
    grid=major,
    grid style={gray!25,dashed},
    legend pos=north east,
    thick
]
\addplot[blue,smooth] table[x=Step,y=Value,col sep=comma]{./table/data_tokens.csv};
\end{axis}
\end{tikzpicture}
\caption{Changes in token consumption during training}
\label{fig:tokens}
\end{figure}

The iteration count rose gradually from 0.0059 at step 1, ultimately reaching a full count of 3 at step 510. Overall, this metric exhibited linear growth relative to the number of training steps, indicating that the training process proceeded in an orderly manner strictly in accordance with the preset number of epochs—and that task execution was free of interruptions or anomalous jumps.

The token consumption during training generally increases linearly show in Figure \ref{fig:tokens}.The combination of hyperparameters selected for this experiment ensured a smooth model fine-tuning process throughout, resulting in ideal training convergence and the absence of anomalous phenomena such as overfitting, underfitting, or training oscillations.

\subsection{RQ2: How effective is our tool DistillGuard in detection on the dataset?}

\begin{table}[]
\large
\centering
\caption{Results of Effectiveness Evaluation}
\label{tab:effectiveness}
\begin{tabular}{@{}cccccc@{}}
\toprule
\textbf{Type} & \textbf{Tool} & \textbf{Acc.} & \textbf{Prec.} & \textbf{Rec.} & \textbf{F1.} \\ \midrule
Static & OSSGadget & 59.6 & 49.7  & \textbf{89.0}  & 63.8   \\
ML & Cerebro       & 83.0 & 98.0  & 58.8           & 73.5   \\
ML & Maltracker    & 86.8 & 87.0  & 78.9           & 82.7   \\
LLM & SocketAI     & 77.0 & 96.1  & 44.4           & 60.7   \\

\midrule
LLM & DistillGuard & \textbf{95.3} & \textbf{99.4} & 88.9  & \textbf{93.8}   \\ 
\bottomrule
\end{tabular}%
\end{table}

In the quantitative performance evaluation shown in Table \ref{tab:effectiveness}.Traditional static analysis tools, exemplified by OSSGadget \footnote{\url{https://github.com/microsoft/OSSGadget}}, rely on rule matching between sensitive APIs and keywords for their core detection logic. While OSSGadget boasts a recall rate of 89.05\%, covering the vast majority of malicious samples, it suffers from severe false positives: a precision of only 49.72\%, an overall accuracy of less than 60\%, and an F1 score of only 63.81\%, placing it in the lower tier among all tools. A deeper analysis of its detection logic reveals that OSSGadget indiscriminately flags all API calls and keyword matches related to sensitive permissions and high-risk operations, regardless of whether they appear in executable code or possess genuine malicious execution logic. This structural flaw results in an extremely low signal-to-noise ratio, making it impossible to fundamentally address the false positive problem through rule updates and hindering its independent use in real-world scenarios.

Machine learning tools, such as Cerebro \cite{zhang2025killing} and Maltracker \cite{yu2024maltracker}, employ a core logic based on training classification models using manually designed code features to automate malicious packet detection. These tools generally face a trade-off between precision and recall, with overall performance significantly lower than that of DistillGuard proposed in this paper. Cerebro, for example, achieves an ultra-high precision of 98.00\%, but its recall is only 58.80\%, and its F1 score is 73.50\%, representing a typical design that sacrifices recall for high precision. Its model learns an extremely narrow decision boundary, only able to label malicious patterns that highly match the training samples, exhibiting very poor generalization ability against novel and mutated malicious packets, resulting in over 40\% of malicious samples being missed.

Maltracker is the highest-performing traditional machine learning tool, achieving 86.86\% accuracy, 87.00\% precision, and 78.95\% recall, with an F1 score of 82.78\%, demonstrating a relatively balanced performance between precision and recall. However, its core bottleneck lies in the quality limitations of manually designed features: traditional machine learning models heavily rely on manually extracted security-oriented features, failing to fully exploit deep semantics, complex execution logic, and hidden malicious behaviors in the code. This significantly reduces its ability to detect obfuscated and transformed malicious packets, and its overall performance still lags behind DistillGuard by more than 11 percentage points in F1 score.

SocketAI \cite{zahan2025leveraging}, as a mainstream LLM detection solution, only achieved 77.04\% accuracy, 96.10\% precision, and 44.40\% recall, with an F1 score of only 60.74\%, even lower than traditional static analysis and machine learning tools, making it unsuitable as a standalone detection tool. The core problem lies in structural flaws in the process design: First, the detection process includes a "critical report" stage, explicitly requiring the large language model to question identified malicious behavior claims, systematically lowering the classification level of otherwise correct malicious packets. Second, the large language model's scoring strategy is overly conservative, placing the classification probability of a large number of genuine malicious packets below the 0.5 threshold, leading to serious missed detections. Third, the input limit of only 20 files per packet causes malicious payloads in deep directory structures to be completely missed, making it impossible to fully analyze the overall behavior of the packet.

Our proposed DistillGuard, a novel LLM detection scheme based on static graph analysis and large model distillation, comprehensively surpasses SocketAI and traditional tools in core metrics: accuracy reaches 95.31\%, precision 99.38\%, recall 88.82\%, and F1 score 93.81\%. Among them, the recall rate is more than twice that of SocketAI, and the F1 score is improved by more than 33 percentage points; at the same time, the F1 score far exceeds the traditional machine learning best solution Maltracker by 11 percentage points, achieving a dual breakthrough in detection accuracy and generalization ability.

\subsection{RQ3: Ablation Study.}

\begin{table}[]
\large
\centering
\caption{Results of Ablation Study}
\label{tab:ablation_study}
\begin{tabular}{@{}c|cccc@{}}
\toprule
 \textbf{Method} & \textbf{Acc.} & \textbf{Prec.} & \textbf{Rec.} & \textbf{F1.} \\ \midrule
Full & \textbf{95.3} & \textbf{99.4} & \textbf{88.9}  & \textbf{93.8}   \\
 \midrule
 (w/o) Pkg Parser   & 91.8 & 99.0  & 80.2           & 88.8   \\
 (w/o) Dep Analyzer & 93.9 & 99.0  & 86.2           & 92.3   \\
 (w/o) Call Graph   & 92.8 & 98.4  & 83.0           & 90.1   \\
 
\bottomrule
\end{tabular}%
\end{table}

To systematically verify the effectiveness and necessity of the three static analysis modules in our tool and investigate the individual contribution of each fine-grained feature to malicious NPM package detection performance, we conduct a set of controlled ablation experiments shown in Table \ref{tab:ablation_study}. All ablation studies maintain identical testing environments and evaluation metrics, with only one static analysis branch removed at a time to ensure fair and credible comparisons. We use False Negatives and False Positives as core error indicators, and compute accuracy, precision, recall, and F1-score. The results are shown in the table.

Without Package.json Parser:The model loses the ability to detect malicious installation hooks such as preinstall and postinstall, leading to a significant increase in FN while FP only rises slightly. The corresponding metrics are accuracy 91.85\%, precision 99.07\%, recall 80.25\%, and F1 88.86\%. This result indicates that configuration-level risk features are critical for identifying automatically triggered malicious behaviors. Without them, the model struggles to capture installation-stage attacks, resulting in an obvious drop in recall.

Without Dependency Analyzer: The model loses perception of high-risk system modules and cannot identify sensitive dependencies such as *os, dns, https, fs* in advance. This causes a mild increase in missed malicious samples and a slight rise in misjudgments of benign packages with normal dependency usage. The corresponding metrics are accuracy 93.97\%, precision 99.02\%, recall 86.25\%, and F1 92.03\%. This shows that dependency features help distinguish benign module usage from malicious exploitation, though their overall impact is moderate.

Without Call Graph Analyzer: The model can no longer understand code execution paths, API call chains, or multi-layer conditional evasion logic. Its ability to detect complex, concealed, and chained malicious behaviors is severely weakened, leading to a clear rise in FN. Meanwhile, FP also increases due to the lack of behavioral path constraints. This result validates that the call graph is the core feature of our approach: it organizes scattered code snippets into interpretable attack behaviors and serves as a key component for improving model generalization and recall.

The full model retains all intermediate features from the three static analysis modules, and performs knowledge distillation and model fine-tuning by integrating configuration risks, dependency risks, and code behavioral logic features. The ablation results demonstrate that removing any single static analysis module leads to measurable performance degradation. This confirms that all three modules are essential components of DistillGuard and provide complementary features to one another.

\subsection{RQ4: Empirical Study of Malicious Behaviors in NPM Packages.}

\begin{table}[htbp]
\normalsize
\centering
\caption{Statistics of Malicious Code Execution Phase}
\label{tab:execution_phase}
\begin{tabular}{@{}c|c@{}}
\toprule
\textbf{Execution Phase} & \textbf{Ratio (\%)} \\
\midrule
Install-Time Attack       & 87.98 \\
Import-Time Attack       & 7.08  \\
Run-Time Attack      & 4.94 \\
\bottomrule
\end{tabular}
\end{table}

We classify malicious npm packages by the execution phase at which their malicious logic is triggered. The results show in Table \ref{tab:execution_phase} that Install-Time Attack is overwhelmingly the most common vector, accounting for nearly 88\% of all malicious cases. This demonstrates that adversaries predominantly abuse npm’s lifecycle hooks such as preinstall and postinstall to execute malicious code stealthily before users run the actual package, making such attacks particularly difficult to detect through conventional runtime monitoring. In contrast, Import-Time Attack and Run-Time Attack represent considerably smaller proportions, at approximately 7\% and 5\%, respectively. Import-time attacks are triggered when the malicious module is loaded, while run-time attacks are activated only when specific functions are invoked. The clear dominance of install-time attacks highlights a critical weakness in current supply chain defenses, as most detection and protection mechanisms focus on runtime behavior rather than installation-time execution.

\begin{table}[htbp]
\normalsize
\centering
\caption{Statistics of Malicious Behavior Types}
\label{tab:maltypes_ratio}
\begin{tabular}{@{}c|c@{}}
\toprule
\textbf{Malicious Types} & \textbf{Ratio (\%)} \\
\midrule
Info-Exfiltration       & 81.51 \\
Command-Execution       & 8.41  \\
Malicious-Download      & 5.52  \\
Credential-Theft        & 2.26  \\
Code-Obfuscation        & 1.17  \\
DNS-Tunneling           & 0.52  \\
Resource-Abuse (Mining) & 0.42  \\
Persistence             & 0.17  \\
\bottomrule
\end{tabular}
\end{table}

We categorize all detected malicious npm packages into eight fine-grained attack types based on their malicious behaviors. As shown in Table \ref{tab:maltypes_ratio}, Info-Exfiltration is the dominant malicious behavior, accounting for 81.51\% of all malicious packages, which indicates that most supply chain threats focus on stealthily collecting host information and exfiltrating it to attacker-controlled servers. The second most prevalent behavior is Command-Execution, representing 8.41\% of the cases, demonstrating attackers’ intent to directly execute system commands and gain execution control over victim machines. Malicious-Download follows as the third common type, making up 5.52\% of malicious packages, which typically fetch and execute remote payloads to evade static detection. In contrast, Credential-Theft, Code-Obfuscation, DNS-Tunneling, Resource-Abuse, and Persistence account for much smaller proportions, yet still pose significant risks including credential leakage, covert communication, cryptocurrency mining, and long-term system compromise.

\begin{table*}[htbp]
\small
\centering
\caption{Malicious API attack chains and common required dependencies}
\label{tab:attack_chain}
\begin{tabular}{@{}p{4cm}|p{7.5cm}|p{5cm}@{}}
\toprule
\textbf{Malicious Types} & \textbf{Typical Malicious API Attack Chain} & \textbf{Dependency}\\
\midrule
Info-Exfiltration       & os.homedir/os.hostname/os.userInfo →  
\newline dns.getServers → https.request → req.write/req.end
  & os, dns, querystring, https \\
\midrule
Command-Execution       & child\_process.exec/execSync/spawn → \newline net.Socket.connect → shell spawn/pipe  & child\_process, net, fs, path \\
\midrule
Malicious-Download      & https/request.get → fs.createWriteStream/ \newline fs.writeFile → child\_process.exec   & fs, https, child\_process, path \\
\midrule
Credential-Theft        & fs.readdir → fs.readFile →   \newline fetch/https.get → axios.post/fetch POST  & fs, node-fetch, axios, https
 \\
\midrule
Code-Obfuscation        & eval → Function → Buffer.from → \newline crypto.createDecipheriv   &      - \\
\midrule
DNS-Tunneling           &os.hostname/os.homedir/process.cwd/ \newline fs.readFileSync → dns.resolve4   & dns, os, fs, path \\
\midrule
Resource-Abuse  & postinstall → child\_process.exec → mining  & child\_process, stratum-client,\newline node-os-utils \\
\midrule
Persistence             & http.createServer → http.Server.listen  &  http, os, fs\\
\bottomrule
\end{tabular}
\end{table*}

Typical malicious API attack chains and their required dependencies are shown in the Table \ref{tab:attack_chain}. Info-Exfiltration is the most pervasive malicious behavior in npm supply chain attacks. Many such attacks are launched at install time by abusing preinstall or postinstall lifecycle hooks to avoid user awareness. The typical API chain first collects system and environment information including home directory, hostname, user info, and DNS servers, then encodes and transmits the collected data to attacker-controlled servers via HTTPS requests. The most frequently imported modules focus on system information acquisition, network communication, and data formatting.

Command-Execution is one of the most hazardous behaviors in npm supply chain attacks, enabling adversaries to gain direct control over compromised hosts. Such attacks occur across both install-time and run-time phases, with most leveraging Node.js’s built-in child\_process module to invoke system shells or arbitrary OS commands. A considerable fraction of Command-Execution attacks further establish reverse shells via network sockets, piping local standard input/output streams to remote attackers for persistent remote control. The most widely used modules focus on process spawning, network communication, and system interaction.

Malicious-Download represents a typical payload delivery behavior in npm supply chain attacks. Such attacks predominantly occur at the install-time phase by abusing npm lifecycle hooks. The typical API call chain involves first fetching remote malicious payloads through HTTPS or HTTP requests, then writing the downloaded content to the local filesystem via file system APIs, and finally executing the dropped payloads using child process mechanisms to achieve arbitrary code execution. 

Credential-theft attacks typically abuse the file system (fs) to read local application storage. A typical example is extracting sensitive tokens through pattern matching and verifying the stolen credentials through the Discord official API, using network libraries such as node-fetch and axios to steal data via Discord Webhook.

Other types of npm malicious code attacks such as Code-Obfuscation DNS-Tunneling Resource-Abuse Persistence are less common, but we still provide typical API attack chains and commonly used malicious dependencies, which still require more attention and defense.

\section{Related Work} \label{sec:related work}


Software supply chain security primarily concerns malicious package detection and registry security \cite{huang2025profmal} \cite{simsek2025pocgen} \cite{zheng2024towards} \cite{yue2024darkfleece} across mainstream platforms, including npm, PyPI, and Docker Hub. Existing malicious package detection methods fall into three categories: rule-based, machine learning-based, and LLM-based approaches. Rule-based methods adopt static analysis and heuristic rules to detect suspicious features such as sensitive API calls and metadata anomalies, represented by GUARDDOG and OSSGadget. Nevertheless, they rely heavily on manual feature engineering, fail to adapt to evolving attack patterns, and suffer from high false positive rates. Machine learning methods utilize handcrafted package features and tree-based models (e.g., lightgbm \cite{ke2017lightgbm}, XGBoost \cite{freund1997decision}) for detection. State-of-the-art methods like AMALFI  \cite{sejfia2022practical} and CEREBRO \cite{zhang2025killing} learn from package metadata, API sequences, and code behavioral semantics, but their performance is limited by feature completeness and vulnerable to obfuscated malicious code. Recently, LLMs have become prevalent for code security analysis. SocketAI  \cite{zahan2025leveraging} and SPIDERSCAN \cite{huang2024spiderscan} exploit iterative prompting and graph-assisted semantic matching to mine malicious code behavior and context. Though effective in semantic comprehension, LLMs are hindered by inherent hallucinations and high computational costs. Hybrid static-dynamic strategies are further proposed to boost detection robustness. DONAPI \cite{huang2024donapi} identifies obfuscated packages via code dependency reconstruction, while MALGUARD \cite{gao2025malguard} combines graph analysis and interpretable algorithms for explainable detection, yet dynamic analysis introduces unavoidable performance overhead.

Apart from package detection, extensive research targets inherent security vulnerabilities of public software and container registries \cite{guo2023empirical} \cite{wu2025exposing}. Typical supply chain threats against npm and PyPI include typosquatting, dependency obfuscation, and account hijacking \cite{sun20241+} \cite{ladisa2023feasibility}, as demonstrated by the classic node-ipc attack incident. Container registries like Docker Hub face diverse risks \cite{wu2025exposing} \cite{wong2023security} \cite{zhu2025doctor}, including outdated components, inherent vulnerabilities, secret leakage, and implanted malicious programs \cite{rosa2023quality}. Prior studies \cite{liu2020understanding} have investigated container quality defects, private secret leakage \cite{dahlmanns2023secrets}, correlations between outdated containers and vulnerable packages, and Dockerfile meta-maintenance schemes \cite{henkel2021shipwright} \cite{zerouali2019relation} \cite{zerouali2019impact}. Overall, most existing detection tools are platform-specific and lack universal generalization across different registries.

\section{Conclusion} \label{sec:conclusion}
This paper presents DistillGuard, a novel NPM malicious package detection framework integrating static graph analysis and LLM knowledge distillation. By combining multi-granularity static features and distilled security domain knowledge with lightweight LoRA fine-tuning, DistillGuard achieves accurate and offline malicious NPM package detection. The experimental results show that DistillGuard substantially outperforms existing state-of-the-art rule, machine learning, and LLM-based detectors, yielding outstanding precision of 99.4\%, recall of 88.9\%, and F1-score of 93.8\%. We further conduct a comprehensive empirical study to quantitatively characterize NPM attack patterns and summarize typical malicious API attack chains. DistillGuard provides an effective, lightweight, data secure detection solution and offers actionable insights for securing the JavaScript open-source supply chain.


\section*{Data Availability}
The code for the DistillGuard tool can be publicly accessed at 
\url{https://doi.org/10.5281/zenodo.20758994}


\bibliographystyle{IEEEtran}
\bibliography{IEEEabrv,references}

\begin{thebibliography}{10}
\providecommand{\url}[1]{#1}
\csname url@samestyle\endcsname
\providecommand{\newblock}{\relax}
\providecommand{\bibinfo}[2]{#2}
\providecommand{\BIBentrySTDinterwordspacing}{\spaceskip=0pt\relax}
\providecommand{\BIBentryALTinterwordstretchfactor}{4}
\providecommand{\BIBentryALTinterwordspacing}{\spaceskip=\fontdimen2\font plus
\BIBentryALTinterwordstretchfactor\fontdimen3\font minus \fontdimen4\font\relax}
\providecommand{\BIBforeignlanguage}[2]{{%
\expandafter\ifx\csname l@#1\endcsname\relax
\typeout{** WARNING: IEEEtran.bst: No hyphenation pattern has been}%
\typeout{** loaded for the language `#1'. Using the pattern for}%
\typeout{** the default language instead.}%
\else
\language=\csname l@#1\endcsname
\fi
#2}}
\providecommand{\BIBdecl}{\relax}
\BIBdecl

\bibitem{cousot1977abstract}
P.~Cousot and R.~Cousot, ``Abstract interpretation: a unified lattice model for static analysis of programs by construction or approximation of fixpoints,'' in \emph{Proceedings of the 4th ACM SIGACT-SIGPLAN symposium on Principles of programming languages}, 1977, pp. 238--252.

\bibitem{tripp2009taj}
O.~Tripp, M.~Pistoia, S.~J. Fink, M.~Sridharan, and O.~Weisman, ``Taj: effective taint analysis of web applications,'' \emph{ACM Sigplan Notices}, vol.~44, no.~6, pp. 87--97, 2009.

\bibitem{chen2016xgboost}
T.~Chen and C.~Guestrin, ``Xgboost: A scalable tree boosting system,'' in \emph{Proceedings of the 22nd acm sigkdd international conference on knowledge discovery and data mining}, 2016, pp. 785--794.

\bibitem{ke2017lightgbm}
G.~Ke, Q.~Meng, T.~Finley, T.~Wang, W.~Chen, W.~Ma, Q.~Ye, and T.-Y. Liu, ``Lightgbm: A highly efficient gradient boosting decision tree,'' \emph{Advances in neural information processing systems}, vol.~30, 2017.

\bibitem{breiman2001random}
L.~Breiman, ``Random forests,'' \emph{Machine learning}, vol.~45, no.~1, pp. 5--32, 2001.

\bibitem{freund1997decision}
Y.~Freund and R.~E. Schapire, ``A decision-theoretic generalization of on-line learning and an application to boosting,'' \emph{Journal of computer and system sciences}, vol.~55, no.~1, pp. 119--139, 1997.

\bibitem{liu2024deepseek}
A.~Liu, B.~Feng, B.~Xue, B.~Wang, B.~Wu, C.~Lu, C.~Zhao, C.~Deng, C.~Zhang, C.~Ruan \emph{et~al.}, ``Deepseek-v3 technical report,'' \emph{arXiv preprint arXiv:2412.19437}, 2024.

\bibitem{team2025kimi}
K.~Team, Y.~Bai, Y.~Bao, Y.~Charles, C.~Chen, G.~Chen, H.~Chen, H.~Chen, J.~Chen, N.~Chen \emph{et~al.}, ``Kimi k2: Open agentic intelligence,'' \emph{arXiv preprint arXiv:2507.20534}, 2025.

\bibitem{du2022glm}
Z.~Du, Y.~Qian, X.~Liu, M.~Ding, J.~Qiu, Z.~Yang, and J.~Tang, ``Glm: General language model pretraining with autoregressive blank infilling,'' in \emph{Proceedings of the 60th Annual Meeting of the Association for Computational Linguistics (Volume 1: Long Papers)}, 2022, pp. 320--335.

\bibitem{hinton2015distilling}
G.~Hinton, O.~Vinyals, and J.~Dean, ``Distilling the knowledge in a neural network,'' \emph{arXiv preprint arXiv:1503.02531}, 2015.

\bibitem{hu2022lora}
E.~J. Hu, Y.~Shen, P.~Wallis, Z.~Allen-Zhu, Y.~Li, S.~Wang, L.~Wang, W.~Chen \emph{et~al.}, ``Lora: Low-rank adaptation of large language models.'' \emph{Iclr}, vol.~1, no.~2, p.~3, 2022.

\bibitem{ouyang2022training}
L.~Ouyang, J.~Wu, X.~Jiang, D.~Almeida, C.~Wainwright, P.~Mishkin, C.~Zhang, S.~Agarwal, K.~Slama, A.~Ray \emph{et~al.}, ``Training language models to follow instructions with human feedback,'' \emph{Advances in neural information processing systems}, vol.~35, pp. 27\,730--27\,744, 2022.

\bibitem{qwen3}
A.~Yang, A.~Li, B.~Yang, B.~Zhang, B.~Hui, B.~Zheng, B.~Yu, C.~Gao, C.~Huang, C.~Lv, C.~Zheng, D.~Liu, F.~Zhou, F.~Huang, F.~Hu, H.~Ge, H.~Wei, H.~Lin, J.~Tang, J.~Yang, J.~Tu, J.~Zhang, J.~Yang, J.~Yang, J.~Zhou, J.~Zhou, J.~Lin, K.~Dang, K.~Bao, K.~Yang, L.~Yu, L.~Deng, M.~Li, M.~Xue, M.~Li, P.~Zhang, P.~Wang, Q.~Zhu, R.~Men, R.~Gao, S.~Liu, S.~Luo, T.~Li, T.~Tang, W.~Yin, X.~Ren, X.~Wang, X.~Zhang, X.~Ren, Y.~Fan, Y.~Su, Y.~Zhang, Y.~Zhang, Y.~Wan, Y.~Liu, Z.~Wang, Z.~Cui, Z.~Zhang, Z.~Zhou, and Z.~Qiu, ``Qwen3 technical report,'' \emph{arXiv preprint arXiv:2505.09388}, 2025.

\bibitem{singh2025openai}
A.~Singh, A.~Fry, A.~Perelman, A.~Tart, A.~Ganesh, A.~El-Kishky, A.~McLaughlin, A.~Low, A.~Ostrow, A.~Ananthram \emph{et~al.}, ``Openai gpt-5 system card,'' \emph{arXiv preprint arXiv:2601.03267}, 2025.

\bibitem{guo2026understanding}
W.~Guo, Z.~Chen, Z.~Xu, C.~Liu, M.~Kang, S.~Song, C.~Liu, Y.~Xu, W.~Sun, and Y.~Liu, ``Understanding npm malicious package detection: A benchmark-driven empirical analysis,'' \emph{arXiv preprint arXiv:2603.27549}, 2026.

\bibitem{Duan2020maloss}
R.~Duan, O.~Alrawi, R.~P. Kasturi, R.~Elder, B.~Saltaformaggio, and W.~Lee, ``Towards measuring supply chain attacks on package managers for interpreted languages,'' in \emph{28th Annual Network and Distributed System Security Symposium, {NDSS} 2021, virtually, February 21-25, 2021}.\hskip 1em plus 0.5em minus 0.4em\relax The Internet Society, 2021.

\bibitem{zhang2025killing}
J.~Zhang, K.~Huang, Y.~Huang, B.~Chen, R.~Wang, C.~Wang, and X.~Peng, ``Killing two birds with one stone: Malicious package detection in npm and pypi using a single model of malicious behavior sequence,'' \emph{ACM Transactions on Software Engineering and Methodology}, vol.~34, no.~4, p. 1–28, 2025.

\bibitem{yu2024maltracker}
Z.~Yu, M.~Wen, X.~Guo, and H.~Jin, ``Maltracker: A fine-grained npm malware tracker copiloted by llm-enhanced dataset,'' in \emph{Proceedings of the 33rd ACM SIGSOFT International Symposium on Software Testing and Analysis}, 2024, pp. 1759--1771.

\bibitem{zahan2025leveraging}
N.~Zahan, P.~Burckhardt, M.~Lysenko, F.~Aboukhadijeh, and L.~Williams, ``Leveraging large language models to detect npm malicious packages,'' in \emph{2025 IEEE/ACM 47th International Conference on Software Engineering (ICSE)}.\hskip 1em plus 0.5em minus 0.4em\relax IEEE, 2025, pp. 2625--2637.

\bibitem{huang2025profmal}
Y.~Huang, W.~Zheng, S.~Wu, B.~Chen, Y.~Lu, Z.~Zhou, Y.~Cao, X.~Li, and X.~Peng, ``Profmal: Detecting malicious npm packages by the synergy between static and dynamic analysis,'' in \emph{2025 40th IEEE/ACM International Conference on Automated Software Engineering (ASE)}.\hskip 1em plus 0.5em minus 0.4em\relax IEEE, 2025, pp. 419--431.

\bibitem{simsek2025pocgen}
D.~Simsek, A.~Eghbali, and M.~Pradel, ``Pocgen: Generating proof-of-concept exploits for vulnerabilities in npm packages,'' \emph{arXiv preprint arXiv:2506.04962}, 2025.

\bibitem{zheng2024towards}
X.~Zheng, C.~Wei, S.~Wang, Y.~Zhao, P.~Gao, Y.~Zhang, K.~Wang, and H.~Wang, ``Towards robust detection of open source software supply chain poisoning attacks in industry environments,'' in \emph{Proceedings of the 39th IEEE/ACM international conference on automated software engineering}, 2024, pp. 1990--2001.

\bibitem{yue2024darkfleece}
C.~Yue, C.~Zhong, K.~Chen, Z.~Zhang, and Y.~Lee, ``$\{$DARKFLEECE$\}$: Probing the dark side of android subscription apps,'' in \emph{33rd USENIX Security Symposium (USENIX Security 24)}, 2024, pp. 1543--1560.

\bibitem{sejfia2022practical}
A.~Sejfia and M.~Sch{\"a}fer, ``Practical automated detection of malicious npm packages,'' in \emph{Proceedings of the 44th international conference on software engineering}, 2022, pp. 1681--1692.

\bibitem{huang2024spiderscan}
Y.~Huang, R.~Wang, W.~Zheng, Z.~Zhou, S.~Wu, S.~Ke, B.~Chen, S.~Gao, and X.~Peng, ``Spiderscan: Practical detection of malicious npm packages based on graph-based behavior modeling and matching,'' in \emph{Proceedings of the 39th IEEE/ACM International Conference on Automated Software Engineering}, 2024, pp. 1146--1158.

\bibitem{huang2024donapi}
C.~Huang, N.~Wang, Z.~Wang, S.~Sun, L.~Li, J.~Chen, Q.~Zhao, J.~Han, Z.~Yang, and L.~Shi, ``$\{$DONAPI$\}$: Malicious $\{$NPM$\}$ packages detector using behavior sequence knowledge mapping,'' in \emph{33rd USENIX Security Symposium (USENIX Security 24)}, 2024, pp. 3765--3782.

\bibitem{gao2025malguard}
X.~Gao, X.~Sun, S.~Cao, K.~Huang, D.~Wu, X.~Liu, X.~Lin, and Y.~Xiang, ``$\{$MalGuard$\}$: Towards $\{$Real-Time$\}$, accurate, and actionable detection of malicious packages in $\{$PyPI$\}$ ecosystem,'' in \emph{34th USENIX Security Symposium (USENIX Security 25)}, 2025, pp. 4741--4758.

\bibitem{guo2023empirical}
W.~Guo, Z.~Xu, C.~Liu, C.~Huang, Y.~Fang, and Y.~Liu, ``An empirical study of malicious code in pypi ecosystem,'' in \emph{2023 38th IEEE/ACM International Conference on Automated Software Engineering (ASE)}.\hskip 1em plus 0.5em minus 0.4em\relax IEEE, 2023, pp. 166--177.

\bibitem{wu2025exposing}
M.~Wu, G.~Hong, W.~Mai, X.~Wu, L.~Zhang, Y.~Pu, H.~Chai, L.~Ying, H.~Duan, and M.~Yang, ``Exposing the hidden layer: Software repositories in the service of seo manipulation,'' in \emph{2025 IEEE/ACM 47th International Conference on Software Engineering (ICSE)}.\hskip 1em plus 0.5em minus 0.4em\relax IEEE Computer Society, 2025, pp. 684--684.

\bibitem{sun20241+}
X.~Sun, X.~Gao, S.~Cao, L.~Bo, X.~Wu, and K.~Huang, ``1+ 1> 2: Integrating deep code behaviors with metadata features for malicious pypi package detection,'' in \emph{Proceedings of the 39th IEEE/ACM international conference on automated software engineering}, 2024, pp. 1159--1170.

\bibitem{ladisa2023feasibility}
P.~Ladisa, S.~E. Ponta, N.~Ronzoni, M.~Martinez, and O.~Barais, ``On the feasibility of cross-language detection of malicious packages in npm and pypi,'' in \emph{Proceedings of the 39th annual computer security applications conference}, 2023, pp. 71--82.

\bibitem{wong2023security}
A.~Y. Wong, E.~G. Chekole, M.~Ochoa, and J.~Zhou, ``On the security of containers: Threat modeling, attack analysis, and mitigation strategies,'' \emph{Computers \& Security}, vol. 128, p. 103140, 2023.

\bibitem{zhu2025doctor}
Z.~Zhu, T.~Chen, C.~Liu, H.~Liu, Q.~Song, Z.~Xu, and Y.~Liu, ``Doctor: Optimizing container rebuild efficiency by instruction re-orchestration,'' \emph{Proceedings of the ACM on Software Engineering}, vol.~2, no. ISSTA, pp. 1--23, 2025.

\bibitem{rosa2023quality}
G.~Rosa, S.~Scalabrino, G.~Bavota, and R.~Oliveto, ``What quality aspects influence the adoption of docker images?'' \emph{ACM Transactions on Software Engineering and Methodology}, vol.~32, no.~6, pp. 1--30, 2023.

\bibitem{liu2020understanding}
P.~Liu, S.~Ji, L.~Fu, K.~Lu, X.~Zhang, W.-H. Lee, T.~Lu, W.~Chen, and R.~Beyah, ``Understanding the security risks of docker hub,'' in \emph{Computer Security--ESORICS 2020: 25th European Symposium on Research in Computer Security, ESORICS 2020, Guildford, UK, September 14--18, 2020, Proceedings, Part I 25}.\hskip 1em plus 0.5em minus 0.4em\relax Springer, 2020, pp. 257--276.

\bibitem{dahlmanns2023secrets}
M.~Dahlmanns, C.~Sander, R.~Decker, K.~Wehrle, J.~Pennekamp, A.~Belova, T.~Bergs, M.~Bodenbenner, A.~B{\"u}hrig-Polaczek, I.~Kunze \emph{et~al.}, ``Secrets revealed in container images: An internet-wide study on occurrence and impact,'' in \emph{ACM Transactions on Internet Technology}, no. 252-266.\hskip 1em plus 0.5em minus 0.4em\relax ACM, 2023, pp. 252--266.

\bibitem{henkel2021shipwright}
J.~Henkel, D.~Silva, L.~Teixeira, M.~d’Amorim, and T.~Reps, ``Shipwright: A human-in-the-loop system for dockerfile repair,'' in \emph{2021 IEEE/ACM 43rd International Conference on Software Engineering (ICSE)}.\hskip 1em plus 0.5em minus 0.4em\relax IEEE, 2021, pp. 1148--1160.

\bibitem{zerouali2019relation}
A.~Zerouali, T.~Mens, G.~Robles, and J.~M. Gonzalez-Barahona, ``On the relation between outdated docker containers, severity vulnerabilities, and bugs,'' in \emph{2019 ieee 26th international conference on software analysis, evolution and reengineering (saner)}.\hskip 1em plus 0.5em minus 0.4em\relax IEEE, 2019, pp. 491--501.

\bibitem{zerouali2019impact}
A.~Zerouali, V.~Cosentino, T.~Mens, G.~Robles, and J.~M. Gonzalez-Barahona, ``On the impact of outdated and vulnerable javascript packages in docker images,'' in \emph{2019 IEEE 26th International Conference on Software Analysis, Evolution and Reengineering (SANER)}.\hskip 1em plus 0.5em minus 0.4em\relax IEEE, 2019, pp. 619--623.

\end{thebibliography}

\end{document}